\documentclass[12pt,a4paper]{article}

\usepackage{amsmath,amsthm,amssymb}
\usepackage{graphics,graphicx}
\usepackage{epsfig}
\usepackage{multicol}
\usepackage{color}
\makeatletter
\@addtoreset{equation}{section}
\makeatother

\begin{document}

\begin{flushright}
QMUL-PH-06-06
\end{flushright}

\begin{center}
\Large\textbf{Non-Abelian (p,q) Strings in the Warped Deformed Conifold.}\\
\vspace{2cm}
\normalsize\textbf{Steven Thomas}\footnote{s.thomas@qmul.ac.uk} \normalsize\textbf{and John Ward}\footnote{j.ward@qmul.ac.uk}\\
\vspace{1cm}
\emph{Department of Physics\\
Queen Mary, University of London\\
Mile End Road, London\\
E1 4NS, U. K}
\end{center}
\vspace{1cm}
\begin{abstract}
We calculate the tension of $(p,q)$-strings in the warped deformed conifold using the non-Abelian DBI action. 
In the large flux limit, we find exact agreement with the recent expression obtained by Firouzjahi, Leblond and 
Henry-Tye up to and including order $1/M^2$ terms if $q$ is also taken to be large. 
Furthermore using the finite $q$ prescription for the symmetrised trace operation we anticipate the most 
general expression for the tension valid for any $(p,q)$. We find that even in this instance, corrections to the
tension scale as $1/M^2$ which is not consistent with simple Casimir scaling.
\end{abstract}
\newpage
\section{Introduction.}
There has been renewed interest in the properties of Fundamental and Dirichlet strings within String Theory. This
has been partly inspired by the recent suggestion that these objects could be cosmic string candidates \cite{cosmicsuperstrings}, however
they have also found prominence due to the gauge theory/ gravity correspondence where the Fundamental strings are expected to terminate
on quarks, and are known by the more generic name of $k$-strings \cite{armoni}.

It has been argued on general grounds that the $F$ and $D$-strings of type IIB string theory should be placed on equal footing. In 
fact this is simply a consequence of $SL(2,Z)$ invariance which implies that both objects are $S$-dual to each other. As a result
it is more natural to consider bound states of such objects, which go by the more colloquial name of $(p,q)$-strings, where
we consider $p F$-strings coupled to $q D$-strings. The tension of such an object in 10D flat space has been calculated to be \cite{tension}
\begin{equation}
T_{(p,q)} = T_F \sqrt{\frac{q^2}{g_s^2}+ p^2},
\end{equation}
which is a manifestly $SL(2,Z)$ invariant expression.
Recent work has aimed to extend this result to non-trivial backgrounds, which have some relevance to gauge theory and
cosmology \cite{bak, curvedspace}. One background in particular, the Klebanov-Strassler throat \cite{conifold}, has come under close scrutiny. The general
idea is that this background is dual to an $\mathcal N =1$ supersymmetric $SU(N+M) \times SU(N)$ confining gauge theory,
where there are $N D3$-branes and $M D5$-branes wrapped on the conifold. In the specific case where $N$ is a multiple of $M$,
the gauge theory cascades down to $SU(M)$ in the deep IR and the $D5$-branes dissolve into $M$ units of RR three-form flux
resulting in the blow up of the $S^3$ within the $T^{1,1}$. This is more commonly referred to as the warped deformed conifold \cite{wdconifold}. Fundamental strings
in this background are dual to a confining string between a quark/anti-quark pair, whilst the $D$-string is dual to an axionic
string \cite{gaugedual}.

From a cosmological perspective we should also be interested in the formation of these $(p,q)$ bound states within this geometry,
as it can be compactified down to four dimensions using closed string fluxes. Typically there will be $D3$/$\bar D3$-branes present in the compactification \cite{kklt}
whose relative motion is a candidate mechanism for inflation \cite{braneinflation}. Once these branes annihilate one another we generally expect the
decay products to be $F$ and $D$-strings, which will energetically prefer to be located at the tip of the warped deformed conifold
and appear as cosmic superstrings in the low energy theory. The range of tensions for these strings is within the current
observational limits and so studying their properties could prove important as a direct verification of string theory.

Previous papers were concerned with calculating the tension spectrum for either the $F$ or the $D$-strings in this background \cite{conifoldstrings}, however
recently this was extended to the $(p,q)$-string bound state \cite{tye}. The idea is to develop a more unified approach that relies on the 
Abelian DBI action for a $D3$-brane, with non-zero electric and magnetic flux on the world-volume. The resulting tension spectrum
was in excellent agreement with the individually calculated tensions. However this prescription is really only strictly valid when the
world-volume fluxes are large in order to minimise non-commutative effects. The result is that we can only trust the solution
for large, integer, values of $p$ and $q$. Interestingly the results of \cite{tye} show that the $F$-strings are charged in $\mathbb Z_M$, whilst the
$D$-strings are charged in $\mathbb Z$. Upon compactification we should expect that this latter condition reduces to
$\mathbb Z_K$ due to the breaking of the global axionic $U(1)$ symmetry, where $K$ is a measure of the bulk flux over the dual cycle of the
manifold.

A different strategy, first attempted by Witten \cite{witten}, is to consider a non-Abelian description of
the problem, where the $F$-strings appear as electric flux on the world-volumes of coincident $D 1$-branes.
This also accounts for the fact that the $D$-strings are charged only under $\mathbb Z$.
In terms of an effective low energy theory we can use the Myers action for coincident $D$-branes \cite{myers, tseytlin}. However we should note that this action
does not agree with string calculations at higher orders, but to leading order provides a very good description. An additional
problem is that the action is typically only calculable in the limit of a large number of coincident branes. However,
a recent proposal has shown how to compute the action for a finite number of branes \cite{finiten}. This is phenomenologically more interesting 
both from the gauge theory side and the cosmology side. In the cosmology case, in particular, we do not expect to see a large number
of cosmic strings in the universe \footnote{In fact we have yet to identify any cosmic strings.} as they provide a negligible
contribution to the Cosmic Microwave Background (see \cite{cosmicstrings} for more comprehensive details). However they are expected to exist as by-products of symmetry breaking. 
Consequently we must ensure our string theory mechanisms for inflation do not over-produce these objects. This is where we
require a description of the string tension for a \emph{finite} number of these strings.

As a first step toward this goal we will initially concentrate on the non-compact case of the warped deformed conifold.
Using the Myers action in curved space \cite{myers, dynamics} we should aim to reproduce the tension spectrum for the $(p,q)$-string initially in the large $q$ limit in order to compare with \cite{tye}, but
we can also find the most general expression for the tension when $q$ is small using the techniques developed in \cite{finiten}. 
In the next section we will review the background solution for the conifold and briefly recapitulate the results obtained in \cite{tye} in the Abelian formulation. In section 3 we will introduce the Myers action and show how we recover the Abelian result in the limit of large flux before proceeding to the case of finite $q$.
Finally we will close with some comments and suggestions for further work.
\section{$(p,q)$ strings in the Warped Deformed Conifold.}
In this section we will review the warped deformed conifold solution \cite{wdconifold} and the construction of the Abelian theory which 
yields the warped $(p,q)$-string tension \cite{tye}. The conifold solution has a Ricci-flat metric which implies that the
base must be a Sasaaki-Einstein manifold. The simplest such space preserving $\mathcal N=1$ supersymmetry is $T^{1,1}$
which is topologically equivalent to an $S^2$ fibration over the three-sphere. We can easily see this from the conifold
metric which is given by
\begin{equation}
d{s_6}^2 = dr^2 + \frac{r^2}{9} \left(d\psi + \sum_{i=1}^2 \cos \theta_i d\phi_i\right)^2 + \frac{r^2}{6} \sum_{i=1}^2 \left( d\theta_i^2 + \sin^2 \theta_i d\phi_i^2\right),
\end{equation}
which clearly implies that $T^{1,1} = (SU(2) \times SU(2))/U(1) \sim S^3 \times S^2$.
However we are interested in the warped deformed conifold which arises once we turn on fluxes through the manifold.
The quantisation condition for the R-R and NS-NS fluxes through the three cycles is given by
\begin{equation}
\frac{1}{4\pi \alpha'} \int_A F_{3} = M, \hspace{0.5cm} \frac{1}{4\pi^2 \alpha'} \int_B H_3 = -K,
\end{equation}
where $M, K \in \mathbf{Z}$ and $A, B$ are dual cycles which are chosen to have an intersection number of one.
Typically the cone is defined by the following expression in $\mathbf {C}^4$, namely $\sum_{i=1}^4 \omega_i^2 =0$,
however we can remove the singular point by deforming the solution as follows
\begin{equation}
\sum_{i=1}^4 \omega_i^2 = \epsilon^2, \hspace{1cm} \epsilon \in \mathbf{R}.
\end{equation}
A convenient representation for the deformed conifold metric is given by introducing a basis of one-forms \cite{conifold} and can be written as
\begin{equation}
d{s_6}^2 = \frac{\epsilon^{4/3}K(\tau)}{2}\left\lbrack \frac{d\tau^2+g_5^2}{3K^3(\tau)} + {\rm cosh}^2(\frac{\tau}{2})[g_3^2+g_4^2] 
+ {\rm sinh}^2 \left(\frac{\tau}{2} \right)[g_1^2+g_2^2] \right\rbrack,
\end{equation}
where the overall warp factor is given by
\begin{equation}
K(\tau) = \frac{(\sinh (2\tau)-2\tau)^{1/3}}{2^{1/3} \sinh (\tau)}.
\end{equation}
Note that $\tau$ parameterises the size of the $S^2$ fibration, and as $\tau \to 0$ we are left with the metric for the
three-sphere. Introduction of the background closed string fluxes allows us to write the full ten-dimensional metric
as follows
\begin{equation}
d{s_{10}}^2 = a_0^2 \eta_{ \mu \nu}dx^\mu dx^\nu + g_s M b_0 \alpha' \left(\frac{1}{2}dr^2 + r^2 d\Omega_2^2 + d\Omega_3^2 \right)
\end{equation}
where now $r$ is the radius of the shrinking two-sphere inside the $S^3$, and is parameterised by $\theta, \phi$.
The $RR$ 2-form in this coordinate system can be written as
\begin{equation}
C_2 = M \alpha' \left(\psi-\frac{\sin(2\psi)}{2} \right)\sin(\theta) d\theta d\phi.
\end{equation}
In order to simulate $(p,q)$ strings in the four-dimensional space one must introduce a $D3$-brane into the conifold which has
non-zero magnetic and electric fluxes on its world-volume. For simplicity both these fields are assumed parallel,
and we choose a gauge such that only $F_{23}$ and $F_{01}$ are non-zero.
The brane is extended in two of the non-compact directions and wrapped on the shrinking 2-cycle inside the $S^3$, where the $B$ field vanishes.
After collecting all the terms and integrating over the compact directions, one can obtain the Hamiltonian density for the 
$(p,q)$ bound state via legendre transform \cite{tye}
\begin{equation}\label{eq:habelian}
H = \frac{a_0^2}{\lambda} \sqrt{\frac{q^2}{g_s^2} + \frac{b_0^2M^2}{\pi^2} \sin^4(\psi) + \left\lbrack\frac{M}{\pi}\left(\psi-\frac{\sin(2\psi)}{2} \right)-(p-qC_0) \right\rbrack^2}.
\end{equation}
At this stage we must minimise the energy with respect to the radius of the $S^2$. The result is 
\begin{equation}\label{eq:psimin}
\psi_{min} + \frac{b_0^2-1}{2} \sin(2\psi_{min}) = \frac{(p-qC_0)\pi}{M}.
\end{equation}
However, the $b_0^2-1$ factor is almost zero and so to leading order we can drop this term. 
Then we see that $\psi_{min} \sim \frac{(p-qC_0)\pi}{M} $
which yields a value $r_{min} $ for the $S^2$ radius at the minimum
\begin{equation}\label{eq:abelian_radius}
r_{\rm min} = \sqrt{b_0 g_s M \alpha'} \left| \sin\left(\frac{\pi(p-qC_0)}{M}\right) \right|
\end{equation}

If we now insert this into our expression for the
Hamiltonian we obtain the tension of the $(p,q)$ bound state
\begin{equation}\label{eq:abelian_tension}
H_{\rm min} = \frac{a_0^2}{\lambda} \sqrt{\frac{q^2}{g_s^2} + \left(\frac{b_0M}{\pi} \right)^2 \sin^2 \left(\frac{\pi(p-qC_0)}{M} \right)},
\end{equation}
which is in excellent agreement with the results obtained in \cite{conifoldstrings}, when one takes the limits $p \to 0$ or $q \to 0$ yielding the tension spectrum
for the $D$ and $F$-strings respectively. If we take the large flux limit then we can expand the
final term in the square root to the next to leading order in powers of $1/M$. This yields the following expression
\begin{equation}\label{eq:abelian_expansion}
H_{\rm min} \sim \frac{a_0^2}{\lambda} \sqrt{\frac{q^2}{g_s^2}+ b_0^2(p-qC_0)^2 \left(1-\frac{\pi^2(p-qC_0)^2}{3M^2}+ \frac{2\pi^4(p-qC_0)^4}{45M^4}-\ldots \right)},
\end{equation}
where we recall that $b_0 \sim 1$. Thus we see that in the large $M$ limit one recovers the expected result for the $(p,q)$ string tension in a 
non-trivial background. It is worth mentioning that if we take the limit $q \to 0$, leaving us with only the fundamental string contribution,
that the corrections to the tension scale as $1/M^2$. This was first noted by Douglas and Shenker \cite{gaugedual} and is different
to the $1/M$ correction that arises due to Casimir scaling \cite{armoni}.
Note that in this limit the minimal radius, $r_{\rm min}$, can be approximated by the following
\begin{equation}
r_{\rm min} \sim \sqrt{b_0 g_s  \alpha'} \frac{\pi(p-qC_0)}{\sqrt{M}}\left|\left \lbrace 1-\frac{\pi^2 (p-qC_0)^2}{6M^2} + \mathcal{O}(\frac{1}{M^4}) \right \rbrace\right|,
\end{equation}
where the terms inside the brackets are a power series in even powers of $1/M$.
\section{Non-Abelian $(p,q)$ strings.}
In this section we shall examine the problem of computing the $(p,q)$ string tension, from the $D$-string perspective. This means we must construct
a theory of $q$ coincident $D$-strings which have $p$ units of quantised electric flux on their world-volumes.
In order to do this we must use the Myers action \cite{myers} for $q$ coincident $D$-strings. The bosonic part can be written as follows
\begin{equation}
S = -T_1 \int d^{2} \xi STr \left(e^{-\phi} \sqrt{- {\rm det}(\mathbf{P}[E_{ab}+E_{ai}(Q^{-1}-\delta)^{ij}E_{jb}]+\lambda F_{ab})} \sqrt{{\rm det} Q^i_j} \right)
\end{equation}
where we are again pulling back the space-time tensors to the $D$-string world-volumes, this time using covariant derivatives.
The first term in the action can be interpreted as a dynamical piece, whilst the final term acts as a potential for the transverse fields.
Note also that $E_{ab} = G_{ab} + B_{ab}$ is the linear combination
of the metric and Kalb-Ramond field, and $Q^i _j = \delta^i_j + i \lambda [\phi^i, \phi^k]E_{kj}$ where $\phi^i$ are the transverse coordinates to
the $D$-string and have canonical mass dimension. The symmetrised trace implies that we must take the trace over 
all symmetrised pairings of the transverse
scalars, however in the limit when $q$, the number of $D$-strings, is large, this reduces to the more familiar trace operation. For the finite $q$ theory
there has been recent progress in developing a consistent prescription for taking the symmetrised trace \cite{finiten}, although this only works for certain
gauge groups.
There also exists the non-Abelian version of the Chern-Simons action, which couples the background RR fields to the coincident $D$-strings. The
action can be written as follows
\begin{equation}\label{eq:nonabcs}
S = \mu_p \int STr \left(\mathbf{P}\left[e^{i \lambda i_{\phi} i_{\phi}} \sum g_s C_{(n)} e^B \right]e^{\lambda F} \right),
\end{equation}
where supersymmetry dictates that $\mu_1 = T_1$. Note that we are working in the conventions of \cite{myers} in which the 
normalization of the RR $C_{(n)}$ differs from the  canonical one, which accounts for the factor of $g_s $ in (\ref{eq:nonabcs}). 
The action contains the interior derivative that maps $p$-forms to $p-2$ forms,
which  for the RR 2-form $C_{(2)} $ gives
\begin{equation}
i_\phi i_\phi C_{(2)} = \frac{C_{ij}}{2} [\phi^j, \phi^i].
\end{equation}
In the above it should be remembered that the field $C_{ij} $ is a function of all the background coordinates and in particular its dependence
on the non-commuting coordinates $\phi^i $ is obtained through the non-Abelian Taylor series \cite{myers}. 
We must also turn on electric flux on the world-volumes of the $D$-strings, which can be interpreted as dissolving $F$-strings onto 
the branes. For simplicity we will also fix the gauge $A_0 = 0$, which implies that $F_{01} \ne 0 = \varepsilon$. Effectively this means
that the gauge field is proportional to the identity matrix in this picture, breaking the $U(q)$ symmetry group of the coincident
branes down to $SU(q) \times U(1)$, where the gauge field now commutes with the $SU(q)$ sector.
\\
\\
If we are very close to the tip of the warped deformed conifold then $B_{(2)}$ vanishes and the dilaton is constant, which simplifies the non-Abelian action.
Furthermore we orient the $D$-strings along two of the Minkowski directions of the non-compact spacetime in order to make contact
with the Abelian theory of the wrapped $D3$-brane. Recall that the $D3$-brane wraps an $S^2$ inside the $S^3$. Since this $S^2$ is thus 'magnetized', 
it suggests that on the non-Abelian side we 
should attempt to describe this wrapped $S^2$ via a fuzzy sphere ansatz for our transverse scalars,
as we know that in the large $q$ limit we should recover the classical two-sphere geometry with $q$ units of magnetic flux.
This is not the same as constructing the dual model to that in \cite{tye}, as in order to do so we would have to consider
a BIon type solution \cite{bions} which blows up into a $D3$-brane wrapped on the two-cycle via the dielectric effect \cite{myers, bachas}.
The non-trivial construction of such a solution is beyond the scope of this note, but would be useful to develop in the future.

Our goal is thus to try and describe, in the non-Abelian theory, a fuzzy 2-sphere embedded not in flat space but in a round $S^3$
geometry, where we capture the essential physics of the solution presented in \cite{tye} but do not construct the dual microscopic
model. Let us begin by only taking the transverse coordinates $\phi^i$ to be non-vanishing in the direction of this $S^3$
whose coordinates we label as $y^a, a=1,2,3$. The metric on this $S^3$ can now be obtained after  
performing a non-Abelian Taylor expansion \cite{myers, verlinde}
\begin{equation}
ds_3^2 = g_{ab} dy^a dy^b, \hspace{1cm} g_{ab}(\phi) \sim g_{ab}(y) + \ldots.
\end{equation}
Since we are not looking for dynamical solutions we can regard the scalar fields as static which simplifies the dynamical portion of the
action. If we calculate the determinant in the potential piece then to leading order we find (using the property that 
$g_{ab}(y)$ is diagonal)
\begin{equation}
{\rm det} Q^i_j = 1-\frac{\lambda^2}{2}[\phi^a, \phi^b][\phi^c, \phi^d] g_{ac}(y) g_{bd}(y) + ..
\end{equation}
and so the DBI contribution to the effective action can be written as
\begin{equation}
S = -T_1 \int d^2\sigma STr \left( a_0^2\sqrt{1-\frac{\lambda^2 \varepsilon^2\mathbf{1}}{a_0^4}} \sqrt{1-\frac{\lambda^2}{2} [\phi^a, \phi^b] [\phi^c, \phi^d] g_{ac}g_{bd}} \right)
\end{equation}
Let us now consider the fuzzy sphere ansatz for the transverse scalars by imposing the following condition
\begin{equation}\label{eq:ansatz}
\phi^a = \hat{R} e^a_{i}\alpha^{i},
\end{equation}
where the $\alpha^{i}$ are the generators of the $SU(2)$ algebra, which is isomorphic to $SO(3)$ and satisfies the 
commutation relation $[\alpha^{i}, \alpha^{j}]=2i \epsilon^{ijk} \alpha^{k}$, and $e^a_i $ are veilbeins on the round three-sphere.
Using this notation the indices $i,j$ label coordinates in the tangent space to this $S^3$.
As in \cite{myers} we will take these generators to be in the $q$ dimensional irreducible representation in order for them
to yield the lowest energy configuration. If we now impose this ansatz on our fields in the action we find
\begin{equation}\label{eq:action1}
S=-T_1 \int d^2\sigma STr \left( a_0^2 \sqrt{1-\frac{\lambda^2 \varepsilon^2}{a_0^4}} \sqrt{1+4 \lambda^2 \hat{R}^4 \hat{C}} \right),
\end{equation}
where $\hat C$ is the usual quadratic Casimir of the representation given by $\hat{C}I_q = \alpha^i \alpha^j \delta_{ij}$, where $I_q$ is the rank $q$ identity matrix. 
It follows from our choice of ansatz in (\ref{eq:ansatz}) there is no explicit dependence of the metric $g_{ab}(y) $ in the above action. 
With the inclusion of the $S^3$ veilbeins in the fuzzy sphere ansatz (\ref{eq:ansatz}), the $SU(2) $ matrices $\alpha^i $ 
arrange themselves into the Casimir invariant $\alpha^i \alpha^i \delta_{ij} $ in the action (\ref{eq:action1}). This feature simplifies the calculation of the 
symmetrised trace both at large $q$ and also for finite $q$ (see next section). It is plausible that there exists a more general choice of ansatz
for the transverse scalars than our proposed solution (\ref{eq:ansatz}), but the resulting $Str$ computation may be rather difficult. 
Another motivation for (\ref{eq:ansatz}) is that
it is easy to see that the equations of motion (assuming constant matrices $\phi^a $ ) are satisfied for the $S^3$ background since the resulting algebraic equations are 
formally equivalent to those obtained in a flat background \cite{myers}, using the ansatz $ \phi^i = \hat{R}\alpha^{i} $.

In the large $q$ limit the symmetrised trace can be approximated
by a trace over the gauge group. We can expand the square roots, take the trace and then re-sum the resultant solution to get a closed form expression for the 
action. Later, when we restrict ourselves to the case of finite $q$ this will change, as we must take the symmetrisation over the scalars into account.

From the expansion of the Chern-Simons action we can see that the leading order non-zero contributions are 
\begin{equation}
S_{cs} = T_1 \int STr \left(\mathbf{P}[C_0 + e^{i\lambda i_\phi i_\phi}C_2] \lambda F \right) + \ldots
\end{equation}
where $C_2$ has only non zero components in the spherical directions. After expanding the action to include the interior derivatives, and performing the
pullback operation we find the action reduces to
\begin{equation}
S_{cs} = T_1 \int d^2 \sigma STr \left(\lambda \varepsilon C_0 - i\lambda^2 \varepsilon \frac{C_{ab}}{2} [\phi^a, \phi^b] \right).
\end{equation}
In order to make any further progress we must Taylor expand the RR two-form which yields a term which will vanish, 
and also a term $ \lambda \partial_c C_{ab} \phi^c$. However under the $STr$ operation  
this terms is proportional to the field strength $F_{abc} $
which gives rise to quantised flux when integrated over $S^3 $. We write $F_{abc} = f \Omega_{abc}$ where 
$\Omega_{abc}$ is the
volume element of $S^3$, and using the flux normalisation condition 
 $\frac{1}{4\pi \alpha'} \int_A F_{3} = M $  we find
\begin{equation}
f = \frac{2}{(b_0 g_s)^{3/2} \sqrt{M \alpha'}}.
\end{equation}
Combining this with the relation $\Omega_{abc} \Omega^{abc}=6$ implies that the 
large $q$ limit of the Chern-Simons action reduces to
\begin{equation}
S_{cs}= T_1 \int d^2 \sigma \left( g_s \lambda q C_0 \varepsilon + \frac{4}{3}\frac{q g_s \hat{C} \lambda^3 \varepsilon 
\hat{R}^3}{(b_0g_s)^{3/2}\sqrt{M\alpha'}}\right).
\end{equation}
\\
\\
We can now construct the canonical momentum density of the electric field by varying the Lagrangian density. As there is no explicit
dependence of the action upon the gauge potential, we expect the resulting quantity to be conserved and also quantised in units
of the string tension.
The resultant displacement field, $p$, is found to be
\begin{equation}
p = T_1qa_0^2 \sqrt{1+4\lambda^2 \hat{C} \hat{R}^4} \frac{\lambda^2 \varepsilon}{a_0^4} \left(1-\frac{\lambda^2 \varepsilon^2}{a_0^4}\right)^{-1/2}
+ T_1 q g_s \lambda C_0 + \frac{4q g_s T_1 \lambda^3 \hat{R}^3 \hat{C}}{3(b_0 g_s)^{3/2} \sqrt{M \alpha'}}.
\end{equation}
Note that we are using the canonical radius $\hat{R}$ in this expression, which is related to the physical radius, $r$, of the fuzzy sphere through the
expression $r^2 = \hat{R}^2 \lambda^2 \hat{C}$. 
If we make this substitution and construct the canonical Hamiltonian density via the Legendre transformation we find
(using the relation $T_1 \lambda = 1/g_s $)
\begin{equation}\label{eq:naenergy}
H = \frac{a_0^2}{\lambda} \sqrt{\frac{q^2}{g_s^2}\left(1+\frac{4r^4}{\lambda^2 \hat{C}} \right)+ \left(p-qC_0-\frac{4qr^3 \sqrt{2 \pi}}{3(b_0 g_s)^{3/2} 
\sqrt{\hat{C} M \lambda^3}}\right)^2}.
\end{equation}
The overall factor multiplying the square root is simply the warped tension of a fundamental string.
At this juncture we should minimise the energy in (\ref{eq:naenergy}) to compare with that predicted in the Abelian theory of the last section. We first
concentrate on the large $M$ approximation, which implies that there is a large flux on the three-sphere. Naively 
one might assume that the energy is minimised when $r = 0$, however we can easily see that this corresponds to a 
saddle point. In fact a quick calculation shows that in this approximation,  the energy is minimised at the following radius
\begin{equation}\label{eq:nonabelian_radius}
r_{\rm min} = \frac{(p-qC_0)}{2 b_0^{3/2}} \sqrt{\frac{2\pi g_s \lambda}{M}}.
\end{equation}
This should be compared to the Abelian result (\ref{eq:abelian_radius}) of the last section.
We see that approximating $b_0 =1$,  to leading order in $1/M$ both expressions for $r_{\rm min} $ are in precise agreement.
Whilst this result is encouraging, what we are really interested in is comparison of the tension of the $(p,q)$ strings in the two formulations.  
Substituting (\ref{eq:nonabelian_radius}) back into (\ref{eq:naenergy}) we find that keeping terms to 
$\mathcal O(1/M^2)$ the energy density at the minimum becomes
\begin{equation}\label{eq:nonabtension2}
H_{\rm min} \sim \frac{a_0^2}{\lambda} \sqrt{\frac{q^2}{g_s^2}\left(1+\frac{(p-qC_0)^4 \pi^2 g_s^2}{b_0^6 M^2 \hat{C}} \right) + 
(p-q C_0 )^2 \left\lbrace 
1-\frac{4(p-qC_0)^2 \pi^2}{3 b_0^6 M^2}\right\rbrace}.
\end{equation}
If we now again work in the approximation where $b_0 = 1$ we can see that the predicted $(p,q)$ string tension  agrees exactly with that predicted in the  
abelian theory (\ref{eq:abelian_expansion}) up to and including terms of $\mathcal O(1/M^2) $.
This result is further strong evidence that the non-Abelian DBI description of $(p,q)$ strings through the fuzzy sphere ansatz
is capturing the correct physics. This is particularly so of the $\mathcal O(1/M^2)$ terms in the tension formula above as these are 
sensitive to the $r^4$ and $r^3 $ terms in the non-Abelian DBI and to our choice of fuzzy sphere ansatz. 

One may wonder if the predicted tension of the $(p,q)$ strings in the non-Abelian formulation agrees (to $\mathcal O(1/M^2) $ ) even if the parameter $b_0 \neq 1$ \footnote{we thank L.Leblond and S.H.Henry-Tye for discussions on this point}. To check this one needs the corresponding expression for the tension in the Abelian formulation, expanded as a power series in $1/M$. This is obtained by first solving the minimization equation (\ref{eq:psimin}) after expanding the 
$\sin(2\psi_{\rm min}) $ term to cubic order. This gives
\begin{equation}
\psi_{\rm min}=\frac{(p-qC_0) \pi}{b_0^2 M} + \frac{2}{3} \frac{(p-qC_0)^3 \pi^3 (b_0^2-1)}{b^8 M^3} + {\mathcal O(1/M^5)} .
\end{equation}

Substituting this value of $\psi_{\rm min} $ into (\ref{eq:habelian}) and expanding in powers of $1/M$ one find precise agreement with the terms arising from a similar expansion of (\ref{eq:nonabtension2}).

A further calculation of the $ \mathcal O(1/M^4) $  in the tension formula shows a discrepancy with the Abelian result. 
The latter predicts corrections $\frac{2}{45M^4}\pi^4 (p-qC_0)^6$  whereas the non-Abelian theory gives
a factor $\frac{4}{9M^4}\pi^4 (p-qC_0)^6$. An investigation of the algebraic structure of sub-leading corrections 
in (\ref{eq:naenergy})
shows that they take the form  $ (\frac{\pi^{2k} (p-qC_0)^{2k+2}}{M^{2k}}) $ for $k=1,2,3 ..$ (taking $b_0 = 1$). 
This is exactly the structure one finds on the abelian side by expanding out the $\sin^2 $ term in 
(\ref{eq:abelian_tension}). 

One may also consider comparing the tension obtained above in the case where the quantised flux  $M$ is not necessarily large.
In this case, $r_{\rm min}$ can be obtained by solving the depressed cubic coming from energy minimization.
 Two of the solutions are imaginary, however the physical solution can be written as follows
\begin{equation}
r_{\rm min} = \left(-\frac{\alpha_0}{2} + \sqrt{\frac{\alpha_1^3}{27} + \frac{\alpha_0^2}{4} }\right)^{1/3} - 
\left(\frac{\alpha_0}{2} + \sqrt{\frac{\alpha_1^3}{27} + \frac{\alpha_0^2}{4}}\right)^{1/3}
\end{equation}
where for large $q$, $ \alpha_0 = -\frac{3}{4} (p-q C_0) \lambda (g_s M \alpha')^{1/3}  $ and $\alpha_1 = \frac{3}{2} M \alpha' g_s $.
It should be noted that to avoid large back reaction corrections to the metric of the $S^3$, $M$ should be taken to be large in order
for us to trust the effective action. Then the perturbative analysis of the string tension in powers of $1/M$  is a good approximation.
\subsection{Tension for Finite $q$.}
The results from the previous subsection were only valid in the large $q$ limit, where we could ignore the additional corrections coming from
the symmetrised trace procedure. If we want to study finite $q$, then this must be taken into account. Fortunately a prescription for doing this has been
recently proposed \cite{finiten}.
Let us write the general form for the Hamiltonian density $H$, where we continue to use the fuzzy $S^2$ ansatz but work in terms of the canonical radius $\hat{R}$
\begin{equation}
H = \frac{a_0^2}{\lambda} \sqrt{\frac{1}{g_s^2}\left(STr \sum_{k=0}^{\infty} \frac{(-1)^k}{k!}\left\lbrack-\frac{1}{2}\right\rbrack_k (4\lambda^2 \hat{C} \hat{R}^4)^k \right)^2+ 
\left(p-\frac{qC_0}{g_s}-\frac{4q \hat{C} \lambda^2 \hat{R}^3}{3 (b_0 g_s)^{3/2} \sqrt{M \alpha'}} \right)^2} \nonumber
\end{equation}
where we have used the following definition
\begin{equation}
\left\lbrack-\frac{1}{2} \right\rbrack_k = \frac{\Gamma(k-1/2)}{\Gamma(-1/2)}.
\end{equation}
Now the symmetrised trace acts on the Casimir of the representation in two different ways, depending on whether the spin representation is odd or even.
There is a simple relationship between the spin and the number of branes, namely $n=2j=q-1$, which will play a role in what follows. The
symmetrised trace acts on the Casimir in the following manner
\begin{eqnarray}
STr [\hat{C}^m] &=& 2(2m+1) \sum_{i=1}^{n/2} (2i)^{2m} \hspace{2.2cm} n = \rm{even}\\
&=& 2(2m+1) \sum_{i=1}^{(n+1)/2} (2i-1)^{2m} \hspace{1cm} n = \rm{odd}. \nonumber
\end{eqnarray}
This prescription implies the following definition for the physical radius of the fuzzy sphere
\begin{equation}
r^2 = \lambda^2 \hat{R}^2 {\rm Lim}_{m \to \infty} \left(\frac{STr \hat{C}^{m+1}}{STr \hat{C}^m} \right) = \lambda^2 \hat{R}^2 n^2,
\end{equation}
where the quadratic Casimir is now $\hat{C}I_q = (q^2-1)I_q = n(n+2)I_{n+1}$ in terms of the spin representation.

We can now consistently take the limit of small $q$ using this prescription. To illustrate this we consider the first
non-trivial solution where there are two coincident $D$-strings. Expansion of the symmetrised trace leads to the following
expression for $H$
\begin{equation}
H = \frac{a_0^2}{\lambda }
\sqrt{\frac{4}{g_s^2}\left(1+\frac{8r^4}{\lambda^2}\right)^2 
\left(1+\frac{4r^4}{\lambda^2}\right)^{-1} + \left(p-2 C_0 - \frac{8 r^3}{(b_0 g_s \lambda)^{3/2}}\sqrt{\frac{2\pi}{M}}\right)^2 }
\end{equation}
where there is a potential sign ambiguity in the $r^3 $ term due to the definition of the physical radius. However we have chosen the
minus sign in order for the solution to agree with that of the large $q$ limit. 
Once again we can search for a minimal radius constraint by considering the large flux limit, which is a useful
simplification. However as there are now only two branes, the backreaction upon the background is more under control.

Writing the full constraint equation for the minimisation of $H$  for the case $q=2$, without demanding that $1/M$ terms are
negligible, we find 
\begin{equation}\label{eq:rmin}
\frac{32r}{g_s^2 \lambda^2} F(r) = \frac{8(p-2C_0)}{(bg_s \lambda )} \sqrt{\frac{2\pi}{M}} G(r) \left(1-\frac{8 r^3}{(p-2C_0)(bg_s\lambda)^{3/2}}\sqrt{\frac{2\pi}{M}} \right)
\end{equation}
where we have introduced the following simplifications
\begin{equation}
F(r) = 1 + \frac{32}{3}\sum_{k=1}^2 k \left(\frac{r^4}{\lambda^2} \right)^k, \hspace{1cm} G(r) = 1 + 8\sum_{k=1}^2 k \left(\frac{r^4}{\lambda^2} \right)^k.
\end{equation}
Clearly eq (\ref{eq:rmin}) is difficult to solve analytically. To simplify the task, we drop all terms of order $1/M$ as in any case they should be insignificant in the large flux solutions we are considering in this note.
There are now two limiting cases of interest for us. The first is when the physical radius
namely $r^4<<\lambda^2$, which allows us to find the solution
\begin{equation}
r = \frac{r'}{2}
\end{equation}
where $r'$ is a shorthand notation for the corresponding minimal radius in the large $q$ case (\ref{eq:nonabelian_radius}) (where we would set q=2). This clearly shows that the
minimum energy configuration occurs at a smaller radius. However we should be careful about interpreting this result, as the Myers action may
not actually be valid in such a limit. Moreover it would also seem to suggest that the $S^2$ embedded within the $S^3$ of the conifold
geometry has shrunk to zero - which would imply a further non-trivial topology change.
The second limit of interest is when the summation is dominated by the $r^8$ terms. Again it is easy to see that the minimal radius
occurs at
\begin{equation}
r = \frac{3 r'}{8}
\end{equation}
which is again smaller than the radius in the large $q$ limit. In fact evaluating the minimum of $H$ for various values of $q$ shows that
this radius is always smaller than the corresponding radius in the large $q$ limit, which is what we would naively expect. 
Fig 1 illustrates this in a plot of the tension $H$ against physical radius $r$ for the three values $q=2,4,10$. Here, we chose 
for convenience $p=100, g_s = 0.1, C_0 = 0.1, M=100, a_0=1$ and work with units where $\alpha' =1 $.
Typically $C_0$ will be small in this solution \cite{tye}, however the results are applicable even when we consider an odd number of
branes.
This shift in radius arises due to the symmetrisation prescription for pairs of fields in the Myers action. Furthermore we see that
the tension at the minimum is smaller for finite $q$, which is phenomenologically interesting from a cosmological perspective
as we can imagine a situation where very few $F$ and $D$ strings are formed at the end of brane/anti-brane annihilation.
The strings will tend to move together to minimise their energy at the tip of the conifold, and for cases where there are only a 
couple of $D$-strings, their respective tensions could easily satisfy the observational bounds \cite{cosmicstrings}.

\begin{figure}[ht]
\centering
\includegraphics[width=10cm]{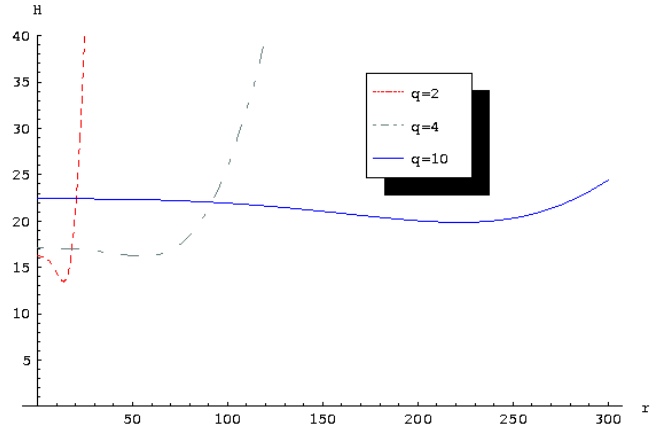}
\caption{string tension $H$ vs $r$ for $q=2,4,10$.}
\end{figure} 
\section{Conclusions.}
In this note we have investigated the $(p,q)$-string tension at the tip of the warped deformed conifold from
the perspective of the non-Abelian DBI action. In the limit of large $p$ and $q$ we recover the same results
as \cite{tye} up to and including $\mathcal O(1/M^2)$ in the large flux expansion. Beyond this order we suspect that higher order corrections to
the DBI action will become relevant, which is why we don't find matching coefficients in the expansion beyond this order. Note that taking $M$ to
be large is necessary in this model in order to neglect back reaction of the fluxes upon the geometry.
If we restrict ourselves to the Fundamental string sector then we obtain the following expansion
\begin{equation}\label{eq:douglas}
H \sim \tilde T_F p \left(1-\frac{p^2\pi^2}{6M^2}+ \ldots \right),
\end{equation}
where we have set $b_0 = 1$ and written $\tilde T_F$ to denote the warped string tension. This is exactly the same
expansion arising from the Douglas-Shenker solution \cite{gaugedual} which supports the conclusion that Casimir scaling
(which yields interaction terms proportional to $1/M$) is not responsible for these corrections \cite{armoni}. Calculating
the higher order terms in the non-Abelian DBI may well provide further agreement with the expected sinusoidal solution,
although this is beyond the scope of the current endeavor. Moreover our current expansion of the action does not
show the expected baryon formation at $p=M$ where the tension vanishes. Again this may change once higher order effects
are taken into consideration.

In addition, we have obtained the general form of the energy density for the case of finite $q$ and shown that the minimisation
radius, and consequently the string tension, increases as we increase $q$. It is more difficult to find an analytic expression for
the string tension in this instance, however we saw that we could make reasonable approximations. The most notable feature of
the finite $q$ result is the form of the $1/M$ expansion. In the limit where $r^4>> \lambda^2, b_0=1$ the tension of the bound state
for $q=2$ cannot be larger than
\begin{equation}
H \sim \tilde T_F \sqrt{(p-2C_0)^2 - \frac{27(p-2C_0)^4 \pi^2}{256M^2} + \ldots}
\end{equation}
where there are higher order interaction terms going as $1/M^4$. If we expand the square root we obtain the same functional form
for the tension as we did in the large $q$ solution - namely corrections proportional to $p^3/M^2$ in the large $M$ limit.
This expansion clearly does not come from Casimir scaling. We cannot say whether it comes from something akin to the
Douglas-Shenker solution without explicit calculation of the higher order terms. However the result is very suggestive,
and shows that the fundamental strings are aware of the finite $q$ effects, which accounts for the numerical discrepancy
with the previous equation (\ref{eq:douglas}).

We can also use this prescription to address the question of Wilson lines in $SU(M)$ gauge theory. A recent paper 
showed how it was possible to calculate the supergravity dual to the Wilson loop in the symmetric representation
using the action for $D3$-branes with electric flux \cite{drukker}. It was further conjectured that the single brane 
gives rise to a single trace operator in the gauge theory. It would be interesting to see if we could use
the finite $q$ prescription for the non-Abelian DBI to calculate multi-trace operators in this $SU(M)$ gauge theory,
especially if we don't have to take $M \to \infty$ in order to obtain a solution (see \cite{diego} for related issues).

The Myers action also explains why the $D$-strings are valued in $\mathbb Z$, at least to leading order. It would certainly
be interesting to investigate the compactified theory, where we expect the $\mathbb Z$ to be broken to $\mathbb Z_K$, as it is
not immediately clear why the number of $D$-strings would be restricted. This is also important from a cosmological perspective,
as we have seen that the $q=2$ solution has a significantly lower tension than the large $q$ results obtained previously.
Therefore it seems preferable for cosmic $(p,q)$ superstrings to be produced in low numbers. We hope to return to this
problem in the future, bearing in mind that there is also a new constraint on the background fluxes with regard to 
string formation after brane/anti-brane annihilation \cite{verlinde}, which implies that $M \ge 12$. Clearly in this
case the $1/M^2$ correction terms could play a more significant role.

A related issue is the dynamics of the bound state in such a geometry. Our solution was purely static and confined to
a specific point in the bulk where the only non-zero fields were the two RR-form fields. A more general problem would be to study the
motion of this bound state in the presence of the non-zero $B$ and four-form RR fields. This would indeed be useful from the
perspective of cosmic strings, and might also have an interesting dual interpretation in the gauge theory.

\begin{center}
\textbf{Acknowledgements.}
\end{center}
It is a pleasure to thank James Bedford and Costis Papageorgakis for their continuing encouragement and comments and Louis Leblond and Henry Tye for useful discussions. JW is supported by a QMUL studentship.
This work is in part supported by the EC Marie Curie research Training Network MRTN-CT-2004-512194.

\end{document}